# Two-terminal spin–orbit torque magnetoresistive random access memory

Noriyuki Sato[1], Fen Xue[1,3], Robert M. White[1,2], Chong Bi[1], and Shan X. Wang[1,2,*]

[1]Stanford University, Department of Electrical Engineering, Stanford, 94305, United States

[2]Stanford University, Department of Material Science and Engineering, Stanford, 94305, United States

[3]Tsinghua University, Department of Electrical Engineering, Beijing 100084, China

[*]sxwang@stanford.edu

## ABSTRACT

Spin-transfer torque magnetoresistive random access memory (STT-MRAM) is an attractive alternative to current random access memory technologies due to its non-volatility, fast operation and high endurance. STT-MRAM does though have limitations including the stochastic nature of the STT-switching and a high critical switching current, which makes it unsuitable for ultrafast operation at nanosecond and sub-nanosecond regimes. Spin–orbit torque (SOT) switching, which relies on the torque generated by an in-plane current, has the potential to overcome these limitations. However, SOT-MRAM cells studied so far use a three-terminal structure in order to apply the in-plane current, which increases the size of the cells. Here we report a two-terminal SOT-MRAM cell based on a CoFeB/MgO magnetic tunnel junction pillar on an ultrathin and narrow Ta underlayer. In this device, an in-plane and out-of-plane current are simultaneously generated upon application of a voltage, and we demonstrate that the switching mechanism is dominated by SOT. We also compare our device to a STT-MRAM cell built with the same architecture and show that critical write current in the SOT-MRAM cell is reduced by more than 70%.

The scaling down of semiconductor devices has improved the speed of computing, but has also increased its power consumption[1]. With conventional computers, dynamic random access memory (DRAM) and static random access memory (SRAM) serve as the main and cache memory, respectively. One possible solution to limit power consumption is to use normally-off computing in which these on-chip volatile memories are replaced with non-volatile memories. For the replacement of DRAM, spin-transfer torque magnetoresistive random access memory (STT-MRAM) is a promising candidate due to its infinite endurance and relatively fast switching times of a few nanoseconds[2–5]. In STT-MRAM devices, however, the initialization of switching is associated with random thermal fluctuations, which makes writing in the sub-nanosecond regime stochastic. As a result, STT-MRAM devices are not practical for replacing SRAM, where fast writing of less than 1 nanosecond is required to keep up with the speed of the CPU[6].

Spin–orbit torques (SOTs), which are generated by an in-plane current and are typically observed in heavy-metal layers[7–11], could provide ultrafast magnetization switching in the sub-nanosecond regime since they do not require an incubation time[12]. SOT-MRAM is also non-volatile and offers high endurance, and could thus potentially be used to replace SRAM and reduce power consumption in computers[13]. However, SOT-MRAM cells developed so far have a three-terminal structure because two isolated contacts[6], with associated interconnects, are required to apply the in-plane current. This inevitably increases the cell size of SOT-MRAM.

In this article, we report a two-terminal SOT-MRAM cell with perpendicular magnetic anisotropy. The cell is made from a CoFeB/MgO magnetic tunnel junction (MTJ) pillar, which has a diameter of 110 nm, on an ultrathin and narrow Ta underlayer. In these devices, an in-plane current and out-of-plane current are simultaneously generated upon application of a voltage. We demonstrate that the switching in the two-terminal MRAM cells is SOT dominant, and that the critical write current in the cells can be reduced by more than 70% compared to the current in equivalent STT-based MRAM cells with 110 nm-diameter MTJs.

**Sample layout**

Figure 1a shows the schematic of the fabricated two-terminal SOT-MRAM cell. The stack structure is Ta(3.8)/CoFeB(1)/MgO(1.2)/CoFeB(1.3)/Ta(0.4)/Co(0.4)/Pd(0.6)/Co(0.4)/Ru(0.85)/Co(0.4)/[Pd(0.6)/Co(0.3)]$_{×3}$/Ru(1.5). The numbers in parentheses indicate the thickness of each layer in the unit of nm. The minimal thickness of the narrow heavy-metal (Ta) underlayer is key to this structure, as suggested in a recently issued patent by IBM[14]. When a voltage is applied, both out-of-plane and in-plane current can be generated. The out-of-plane current induces a STT, while the in-plane current induces a SOT. Our devices are different from previous concepts that combine STT and SOT in a single device, which still require three-terminals[15,16]. As shown below, in our devices the contribution of STT in the switching is much lower than the SOT contribution. The maximum in-plane current density ($J_{in}$) in our device is related to the out-of-plane current density ($J_{out}$) by the following equation[14]:

$$J_{\text{in}} \text{ (max.)} = \frac{\pi d_{\text{MTJ}}^2}{4w_{\text{HM}}(t_{\text{HM}}+t_{\text{FM}})} J_{\text{out}} \tag{1}$$

where $w_{\text{HM}}$ is the width of the heavy-metal underlayer, $t_{\text{HM}}$ is the thickness of the underlayer, $t_{\text{FM}}$ is the thickness of the ferromagnetic layer, and $d_{\text{MTJ}}$ is the diameter of MTJ pillar. Note, the reduction of $t_{\text{HM}}$ cannot simply be used to increase the in-plane current density because the effective spin Hall angle reduces as the heavy-metal underlayer becomes thinner. This reduction can be expressed by an effective spin Hall angle ($\theta_{\text{eff}}$) given by:

$$\theta_{\text{eff}} = \theta[1 - \text{sech}(\frac{t_{\text{HM}}}{\lambda_{\text{sd}}})] \tag{2}$$

where $\theta$ is the spin Hall angle of the heavy-metal underlayer and $\lambda_{\text{sd}}$ is the spin-diffusion length in the heavy-metal layer. To maximize the spin Hall-induced torque, we need to maximize the product of $J_{\text{in}}$ and $\theta_{\text{eff}}$, which are given by the above two equations, respectively. We determined $\lambda_{\text{sd}}$ in the Ta underlayer to be 2.0 nm from a separate experiment (see Supplementary Information) and selected the heavy-metal underlayer thickness of 3.8 nm to maximize the torque induced by an in-plane current.

Figure 1b shows the scanning electron microscope (SEM) image of the MTJ pillar on the Ta underlayer. The width of the Ta underlayer is 220 nm and the diameter of the MTJ pillar is 110 nm (see Methods). Using Eq. 1, the maximum ratio of $J_{\text{in}}$ to $J_{\text{out}}$ of this device is estimated to be 14.3. We measured the resistance versus out-of-plane magnetic field, as shown in Fig. 1c. We observe a sharp change in the resistance, which indicates quasi single-domain switching. At the negative field, the parallel-state resistance increases slightly as the magnitude of the negative magnetic field is increased. This is due to the weak perpendicular magnetic anisotropy of the reference layer (see Supplementary Information). The difference in resistance between the parallel state and the anti-parallel state at the zero magnetic field is only 8% because the measured resistance includes the contribution from both the tunnel junction and the Ta wire. The Ta wire has a larger resistance than the tunnel junction in our devices. In order to measure the resistance of the tunnel barrier alone, we performed a four-terminal measurement and obtained a TMR ratio of 60%.

## Confirmation of SOT-dominant switching

Figure 2 shows the current-induced resistance change under different external fields and reference-layer magnetization states. The current pulse width is set at 10 µs. External magnetic field is applied along current direction (x-axis). Comparing Fig. 2a, b, we note two distinct features that suggest the switching mechanism is SOT-dominant under the above condition: (1) the switching polarity becomes opposite when reversing the applied magnetic field direction. In SOT switching, an in-plane magnetic field breaks the symmetry, and the up or down state becomes the ground state depending on the direction of the applied magnetic fields[8,9]; (2) the critical current density is almost identical for the parallel-to-anti-parallel (P-to-AP) switching and anti-parallel-to-parallel (AP-to-P) switching. This is in sharp contrast to the conventional STT switching, in which the critical current for P-to-AP switching is up to a few times larger than that of AP-to-P switching[2]. These two features clearly suggest that the current-induced magnetization switching mechanism in this device is SOT-dominant. This is further supported by the results to be discussed in the next section when we vary the width of Ta underlayer. Additionally, we note that the critical current density under $H_x$ = 300 Oe is slightly smaller than that for $H_x$ = -300 Oe. If there is only SOT, the critical current density under the same magnitude of the in-plane magnetic field should be identical. In our device, STT is also generated by the out-of-plane current and it should reduce or increase the SOT critical switching current depending on the SOT switching direction. When SOT switching direction is the same as that of STT, the critical switching current should be reduced. Otherwise, the critical switching current should be increased. It appears that the STT assists the switching under the positive magnetic field whereas it slightly increases the switching current under the negative magnetic field.

To further confirm this speculation, we performed the same experiment except that the magnetization direction of the reference layer is reversed so that the effect of STT is reversed. In this case, as expected, STT assists the switching under the negative magnetic field by considering a slight increase of the switching current under the positive magnetic field as shown in Fig. 2c,d. This result also suggests that STT effect is minimal in this device. Thus, we can conclude that the dominant switching mechanism is SOT switching and STT contribution to the critical current is approximately 10%. It is very interesting to compare these experimental results to those theoretical predictions by combining SOT and STT even though the three-terminal structures were still adopted in those works[17]. Lee et al. estimated the ratio of the critical current arising from SOT to STT to be 5.5–18 in a structure similar to our device[17]. This means that in our device, where the ratio of $J_{in}$ to $J_{out}$ is 14.3 at maximum, the contributions from the two effects should be similar. However, our experimental results clearly suggest that the switching mechanism is SOT-dominant. We speculate that this disagreement can be explained by: (1) enhanced effective damping parameter due to the large in-plane current density[18,19], which leads to an increased critical current for the STT-switching mechanism; (2) spin polarization is lower than the typical CoFeB-MgO-based perpendicular MTJ because the perpendicular anisotropy of the reference layer in our device is weak (see Supplementary Information). In the following experiments, we keep the reference-layer magnetization pointing to +z and the in-plane external magnetic field positive so that the SOT switching is assisted by STT and thus the critical current is minimized.

**Reduction of critical current**

Next, we investigate the critical current versus the current pulse width in order to extract the intrinsic critical current $I_{c0}$. In the above discussion, we concluded that the switching mechanism under external magnetic fields is SOT-dominant. Thus, we use the following equation corresponding to the thermal activation model[20] for the SOT critical current to fit the data with pulse length of 10 ns or longer:

$$I_c = I_{c0} \left[ \pi - 2h_x - \sqrt{8\left(\frac{\ln(\tau/\tau_0 \ln(2))}{\Delta} - 1\right) - 4h_x^2 - 4h_x(\pi - 4) + \pi^2} \right] \tag{3}$$

where $h_x = H_x/H_{keff}$, $H_x$ is the external in-plane magnetic field, $H_{keff}$ is the effective perpendicular anisotropy field, $\tau$ is the pulse width, and $\tau_0 = 1$ ns is the inverse of the attempt frequency. The extracted intrinsic critical currents are 778 µA (P to AP) and -767 µA (AP to P), which correspond to the in-plane current density of 73.7 MA/cm$^2$ and -72.6 MA/cm$^2$ under +300 Oe, respectively. The thermal stability ratio Δ of 40.9 is also obtained. In the case of the pure SOT switching mechanism, the intrinsic critical current is given by[17]:

$$I_{c0\_SHE} = \frac{2e}{\hbar} \frac{M_s dA}{\theta} \left( \frac{H_{keff}}{2} - \frac{H_x}{\sqrt{2}} \right) \quad (4)$$

where $A$ is the cross-sectional area for the in-plane current. The θ of the Ta layer is assumed to be 0.15. $H_{keff}$ is extracted from the magnetization curve measurement. The theoretical value of $I_{c0}$ is 3.8 mA, which is approximately five times larger than the experimental result. This difference is too large to be explained by the increase of $\theta_{eff}$ due to the Rashba–Edelstein effect[19]. Moreover, the STT contribution to the critical current was estimated to be only ~ 10% as shown above, which is also too small to explain the difference. Thus, we attribute this difference to the domain wall motion-driven magnetic switching where the switching starts from domain nucleation and is followed by the propagation of domain walls. This is also supported by the linear scaling of the critical current with ~ $1/\tau$[12] for the pulse length less than 10 ns, where the thermal activation does not help the switching. Since the field-like torque in Ta/CoFeB system[11] is sizeable compared to Pt/ferromagnet system[10], it may assist domain nucleation in our devices during the initial switching process[21].

Moreover, the effect of the Ta underlayer width on the critical current is studied. We fabricated a series of samples with different underlayer widths, from 220 nm to 20 µm so that the effect of STT can be evaluated. Figure 3b shows the critical current as a function of the Ta underlayer width. The in-plane magnetic field was positive and the magnetization of the reference layer was pointing in the +z direction in order to get STT-assisted SOT switching. The pulse width was set at 10 ms. When the underlayer width is 20 µm, $J_{in}$ is much smaller than $J_{out}$ based on Eq. 1. Thus, it is safe to assume that the switching is purely due to STT switching. On the other hand, when the underlayer width is 0.22 µm, the switching is dominantly caused by SOT mechanism, as discussed above. As the underlayer width is reduced, the critical current gradually decreases and becomes almost one-third of that for pure STT switching. This result suggests that the switching in the two-terminal SOT-MRAM cells can be more efficient than the STT-MRAM cells by taking advantage of the SOT effect.

## Conclusions

We fabricated two-terminal SOT-MRAM MTJ structures on a thin and narrow Ta underlayer, a configuration provides a large in-plane current density and hence produces a large SOT effect. We showed that it is possible to significantly reduce the critical current in the two-terminal MRAM cell by using SOT switching, and demonstrated magnetization switching with a pulse current as short as 3 ns. Our device offers a number of benefits: good potential scalability due to the two-terminal structure; lower critical current density than a device based purely on STT switching (recent work has also shown that the switching current of conventional two-terminal MTJs can be reduced by introducing SOT[22]); and much faster switching speed since no thermal incubation time is required at the initialization of the switching[12]. The devices do though have some potential drawback. For example, they reduce the apparent tunnel magnetoresistance ratio due to the large resistance of the Ta underlayer.

Furthermore, the devices require an in-plane external field to induce the SOT effect. However, recent reports have shown SOT switching in the absence of a magnetic field[23–25]. Therefore, it should be possible to achieve field-free SOT switching in future two-terminal SOT-MRAM devices by introducing a structural asymmetry or antiferromagnetic coupling. Finally, the difficulty of fabricating the very thin Ta wire is a potential obstacle for the mass production of these devices. Despite these challenges, we believe these two-terminal SOT-MRAM devices are capable of replacing SRAM in cache, and could help lead to the development of normally-off computing technology.

## Methods

**Sample preparation.** We first deposited Ta(3.8)/CoFeB(1)/MgO(1.2)/CoFeB(1.3)/Ta(0.4)/Co(0.4)/Pd(0.6)/Co(0.4)/Ru(0.85)/Co(0.4)/[Pd(0.6)/Co(0. 3)]$_{\times 3}$/Ru(1.5) on thermally oxidized Si wafer by using the AJA DC and RF magnetron sputtering system. The numbers in parentheses are in nm. Then, we annealed the sample at 200 °C in vacuum for one hour. A perpendicular magnetic field of 2 kOe was applied during the annealing process using a high-temperature-tolerant permanent magnet. We performed an electron beam lithography process with an MaN2403 negative-tone resist to pattern the heavy-metal underlayer with the Raith electron beam patterning system. Then, we etched all the layers using Intlvac Ion Beam Mill Etcher, which was followed by the removal of the photoresist using Remover PG in an ultrasonicator. The second electron beam lithography process with MaN2403 resist is then performed to pattern the MTJs. Then, we etched all the layers except the bottom Ta layer—i.e. we etched the CoFeB(1)/MgO(1.2)/CoFeB(1.3)/Ta(0.4)/Co(0.4)/[Pd(0.6)/Co(0.16)]x4/Pd(0.6)/Ru(0.85)/Co(0.5)/[Pd(0.6)/Co(0.16)]x7/Pd(0.6)/Ru(1. 5) layers by using Intlvac Ion Beam Mill Etcher. The thickness of the remaining Ta layer was monitored by measuring the resistance of a dummy Ta wire on the same wafer. The sample was immediately transferred to the vacuum chamber of the Laybold RF magnetron sputtering system to prevent the oxidation of the very thin Ta underlayer. We deposited $Al_2O_3$ (22.5 nm) without breaking the vacuum. The third electron beam lithography process with ZEP520 positive-tone resist was then performed to pattern the first top electrode. A sputter etch of the top 1 nm followed by a deposition of Al (200 nm) was performed with the AJA DC and RF magnetron sputtering system. Lastly, a photolithography with positive-tone Shipley 3612 photoresist was performed to pattern the second top electrode using a Kersuss mask aligner. Another sputter etch of 1 nm followed by a deposition of Al (200 nm) was performed and the fabrication was finalized with a lift-off process.

**Data availability.** The data that support the plots within this paper and other findings of this study are available from the corresponding author upon reasonable request.

**Acknowledgements**

S.X.W. wishes to thank TSMC, Stanford SystemX Alliance, Stanford Center for Magnetic Nanotechnology, and the NSF Center for Energy Efficient Electronics Science (E3S) for financial support. N.S. would like to thank Funai Foundation for Information Technology for the overseas scholarship. This work was supported in part by ASCENT, one of six centers in JUMP, a Semiconductor Research Corporation (SRC) program sponsored by DARPA. The experimental work has benefited from the equipment and tools at Stanford Nanofabrication Facility, Stanford Nano Shared Facilities, and Michigan Lurie Nanofabrication Facility (LNF) which are supported by the National Science Foundation (NSF).

**Author contributions**

N.S., R.M.W., and S.X.W. conceived the experiments, N.S. and F.X. conducted the experiments, and N.S., C.B. and S.X.W. analyzed the results. All authors reviewed the manuscript.

**Competing interests**

The authors declare no competing interests.

**Additional information**

**Supplementary information** is available for this paper at https://doi.org/.

**Reprints and permissions information** is available at www.nature.com/reprints.

**Correspondence and requests for materials** should be addressed to S.X.W.

**Publisher's note:** Springer Nature remains neutral with regard to jurisdictional claims in published maps and institutional affiliations.

**Figure captions**

**Fig. 1| Characterization of two-terminal spin-orbit torque devices. a**, Schematic of the cell of two-terminal spin–orbit torque magnetoresistive random access memory. **b**, Scanning electron microscope image of the MTJ pillar on the Ta underlayer. **c**, Resistance as a function of applied out-of-plane magnetic field.

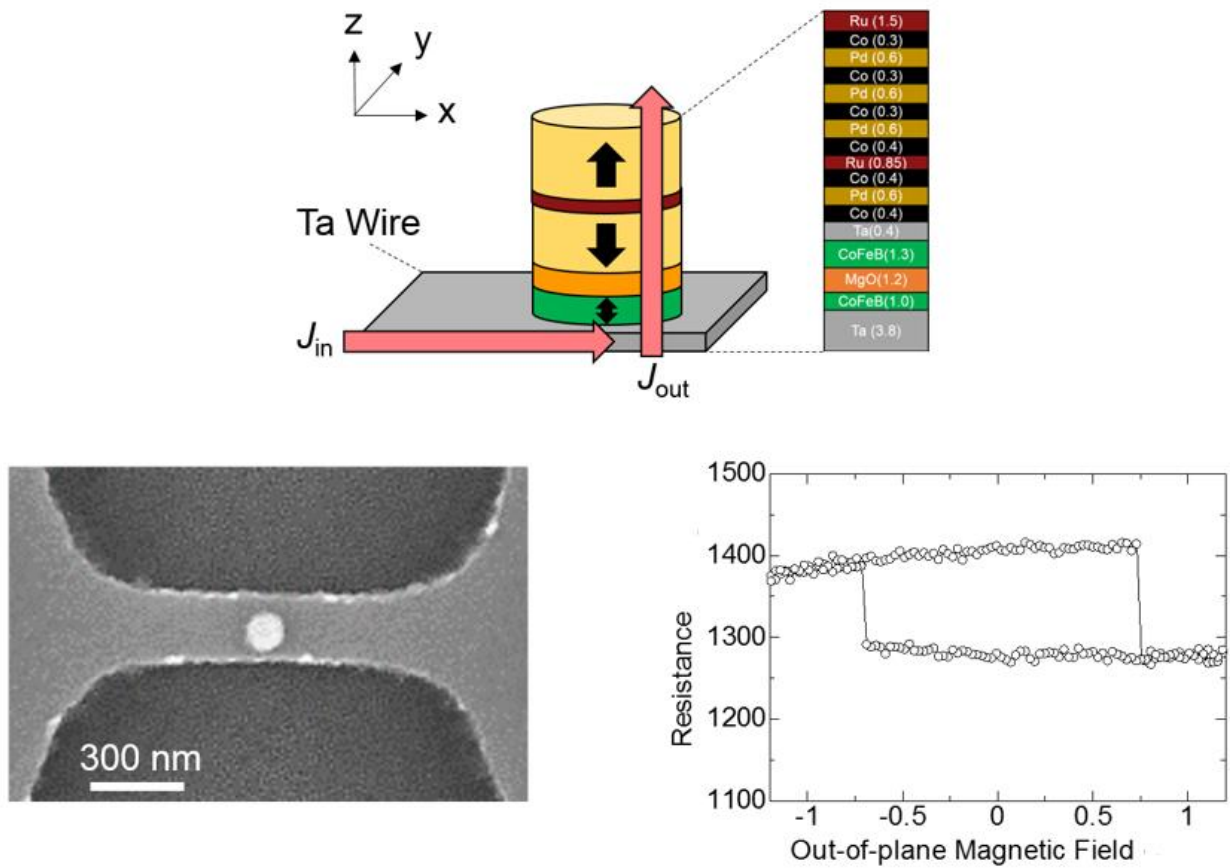

**Fig. 2|Current-induced magnetization switching in two-terminal spin-orbit torque devices. a**, $m_{ref,z} > 0$ and $H_x = 300$ Oe. **b**, $m_{ref,z} > 0$ and $H_x = -300$ Oe. **c**, $m_{ref,z} < 0$ and $H_x = 300$ Oe. **d**, $m_{ref,z} < 0$ and $H_x = -300$ Oe. $m_{ref,z}$ is the magnetization component of the reference layer along $z$ direction. $H_x$ is the applied in-plane field.

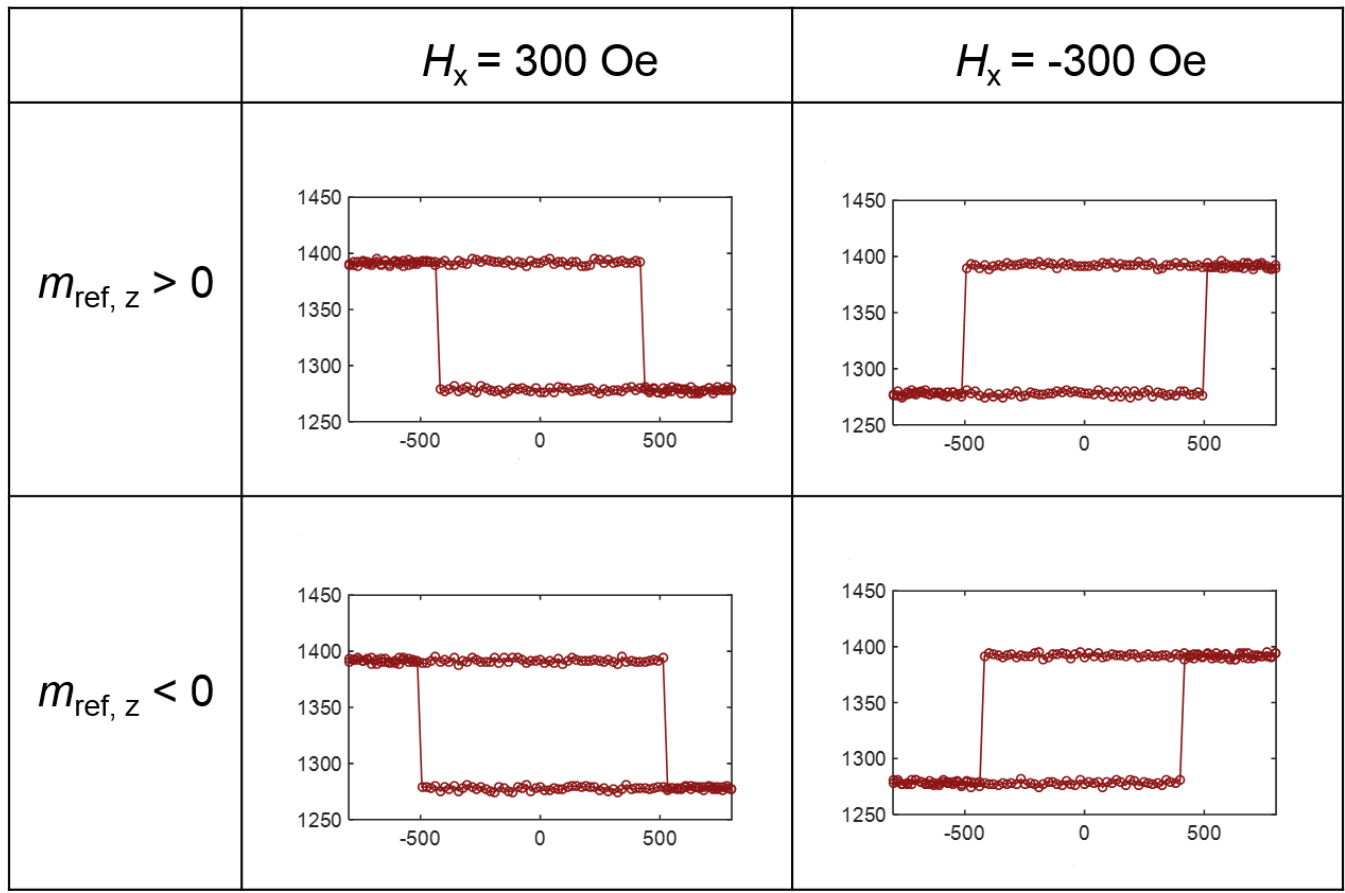

**Fig. 3 | Pulse width and $w_{Ta}$ dependence of two-terminal spin-orbit torque switching. a**, Pulse width dependence of the critical current for the device with 220 nm Ta underlayer. **b**, Ta wire width dependence of the critical current at 10 ms.

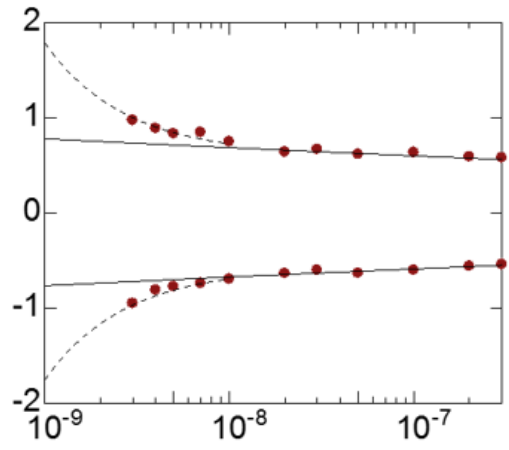 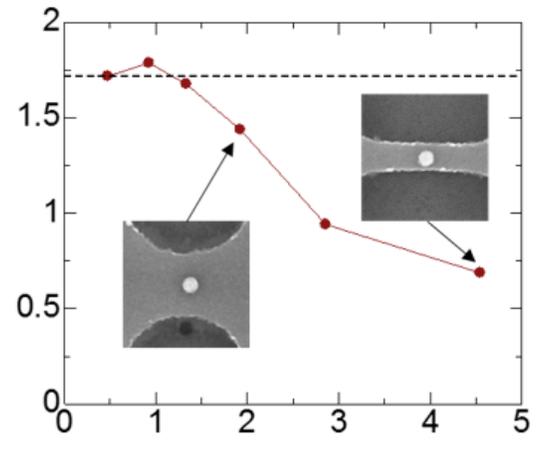

**Table of Contents:**

**Supplementary Note 1: Spin-diffusion length in Ta**

In order to determine the optimum thickness of the Ta underlayer, it is critical to measure the spin-diffusion length in Ta. It is not desirable to take a literature value because the different spin-diffusion lengths in Ta have been observed from different groups. Supplementary Figure 1 shows the longitudinal effective field as a function of the thickness of the bottom Ta layer in Ta/CoFeB(1.2 nm)/MgO(1.5 nm)/Ta(1 nm) multilayer structure. The effective field is measured by low-current induced lock-in technique[1]. The effective field is normalized by that of $t_{Ta}$ = 10 nm. The spin-Hall-effect-induced effective field initially increases sharply with the bottom Ta layer thickness and almost saturates above $t_{Ta}$ = 5 nm. The spin-diffusion length was extracted by the following equation:[2]

$$\frac{\theta_{\text{SH-eff}}}{\theta_{\text{SH}}} \sim \frac{\Delta H_L(t)}{\Delta H_L(t = 10 \text{ nm})} = 1 - \text{sech}\left(\frac{t_{Ta}}{\lambda_{Ta}}\right) \quad (1)$$

where $\theta_{SH}$ is the spin Hall angle, $\Delta H_L$ is the effective field induced by spin-Hall effect, $t_{Ta}$ is the bottom Ta layer thickness and $\lambda_{Ta}$ is the spin-diffusion length in Ta. The fitting result (red solid line) is shown in Supplementary Fig. 1. We found that the spin-diffusion length in Ta is ∼ 2.0 nm for our sample.

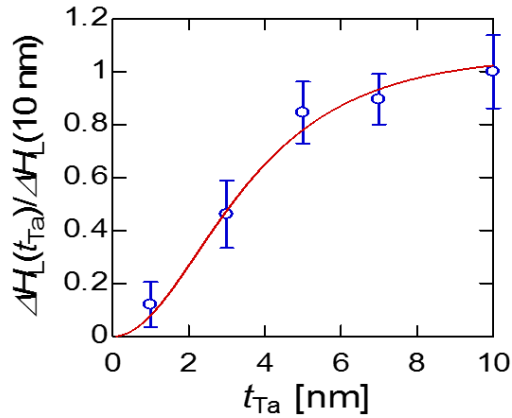

**Supplementary Figure 1|** Effective field induced by spin-Hall effect as a function of bottom Ta layer thickness.

**Supplementary Note 2: Determining Ta Underlayer Thickness**

The device schematic is shown in Supplementary Fig. 2. The thin and narrow heavy-metal underlayer is the key for this structure. Upon the application of current, both out-of-plane and in-plane current will be generated. The former generates a spin-transfer torque which leads to a slower precessional switching. The latter corresponds to a spin-orbit torque that should provide a non-precessional fast switching. The in-plane current density in this device is related to the out-of-current density by the following equation:[3]

$$J_{\text{in}} \text{ (max.)} = \frac{\pi d_{\text{MTJ}}^2}{4w_{\text{HM}}(t_{\text{HM}}+t_{\text{FM}})} J_{\text{out}} \quad (2)$$

where $w$ is the width of the underlayer, $t_{\text{HM}}$ is the thickness of the underlayer, $t_{\text{FM}}$ is the thickness of the ferromagnetic layer and $d_{\text{MTJ}}$ is the diameter of the MTJ. One may think that we should simply reduce the thickness of the underlayer as much as possible to increase the in-plane current density. However, when the thickness of the heavy-metal underlayer is thinner, this effective reduction of the spin-Hall angle must be considered. The reduction can be expressed by effective spin-Hall angle $\theta_{\text{eff}}$ given by:

$$\theta_{\text{eff}} = \theta \left[1 - \text{sech}\left(\frac{t_{\text{HM}}}{\lambda_{\text{sd}}}\right)\right] \quad (3)$$

where $\theta$ is the spin-Hall angle of the heavy-metal underlayer, $t_{\text{HM}}$ is the thickness of the heavy-metal underlayer and $\lambda_{\text{sd}}$ is the spin-diffusion length in the heavy-metal layer. Therefore, the strength of spin-Hall-effect-induced torque can be expressed as:

$$H_s = \frac{\pi d_{\text{MTJ}}^2}{4w(t_{\text{HM}}+t_{\text{FM}})} \frac{\gamma \hbar \theta}{2eM_s t_{\text{FM}}} \left[1 - \text{sech}\left(\frac{t_{\text{HM}}}{\lambda_{\text{sd}}}\right)\right] J_{\text{out}} \quad (4)$$

where $t_{\text{FM}}$ is the thickness of the ferromagnetic free layer, $\gamma$ is the gyromagnetic ratio, $\hbar$ is the reduced Plank constant, $M_s$ is the saturation magnetization of the free layer. Using the above equation, $H_s$ is calculated as a function of heavy-metal underlayer thickness as shown in Supplementary Fig. 3. The data here is normalized to the maximum value. $H_s$ is maximized at different heavy-metal underlayer thicknesses for different spin-diffusion length $\lambda_{\text{ds}}$ of the heavy-metal underlayer. We selected Tantalum as the underlayer material because it has a large spin-Hall effect and Ta/CoFeB/MgO multilayer is known to provide a perpendicular magnetic anisotropy when CoFeB layer is sufficiently thin. In addition, Ta underlayer is an excellent Boron sink owing to the low tantalum boride formation enthalpy[4], which is critical to obtaining a high tunnel magnetoresistance in CoFeB-MgO based magnetic tunnel junction[5]. We extracted the spin diffusion length of Ta underlayer in our Ta/CoFeB/MgO multilayer samples using the low-current induced lock-in technique as described above. At $\lambda_{\text{ds}}$ = 2.0 nm, $H_s$ is maximized when the heavy-metal underlayer thickness is 3.8 nm. Thus, we select 3.8-nm-thick Ta underlayer as the source of spin-Hall effect in our two-terminal device.

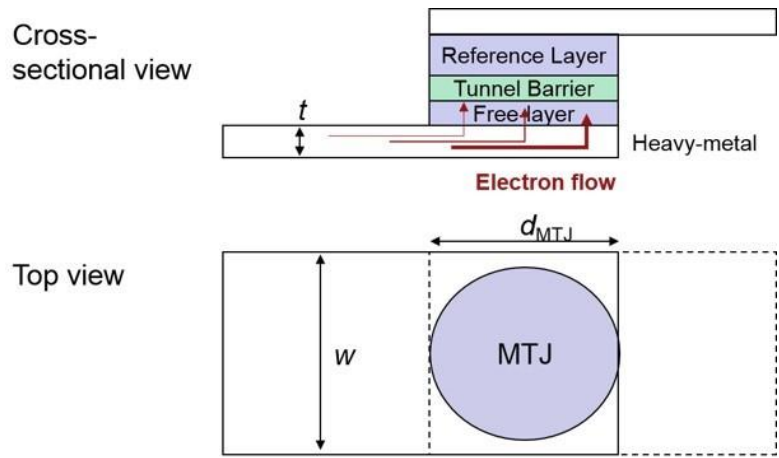

**Supplementary Figure 2| Schematic of the two-terminal device.** $w$ is the width of the underlayer, $t$ is the thickness of the underlayer and $d_{MTJ}$ is the diameter of the MTJ. The thin and narrow heavy-metal underlayer is the key for this structure. Upon the application of current, both out-of-plane and in-plane current will be generated.

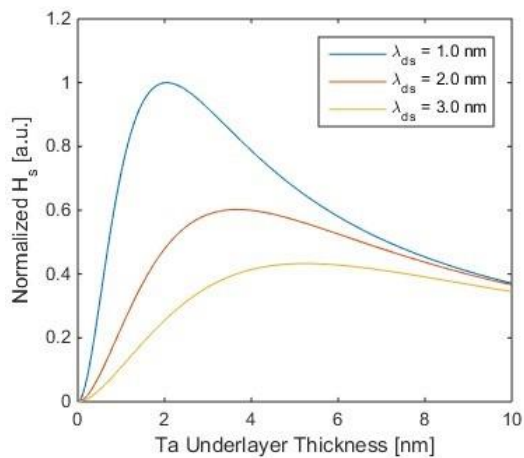

**Supplementary Figure 3| Calculated spin-Hall-induced effective field (normalized by maximum value) as a function of the Ta underlayer thickness with various spin-diffusion length in the Ta layer.**

## Supplementary Note 3: MTJ Stack

Supplementary Figure 4 shows the out-of-plane magnetization curve of a blanket film with the designed stack structure Ta(3.8)/CoFeB(1)/MgO(1.2)/CoFeB(1.3)/Ta(0.4)/Co(0.4)/Pd(0.6)/Co(0.4)/Ru(0.85)/Co(0.4)/[Pd(0.6)/Co(0.3)]x3/Ru(1.5) for the two-terminal device. The blanket film is post annealed at 200 °C for 1 hour. The switching of the CoFeB free layer and the Co(0.4)/[Pd(0.6)/Co(0.3)]x3 pinned layer can be observed clearly. On the other hand, the switching of the reference layer is characterized by a gradual rotation which indicates a weak perpendicular anisotropy[6]. The total thickness of this stack structure is 14.6 nm, which is thinner than typical full MTJ stacks. This allows us to reduce the overetching thickness to less than 1 nm.

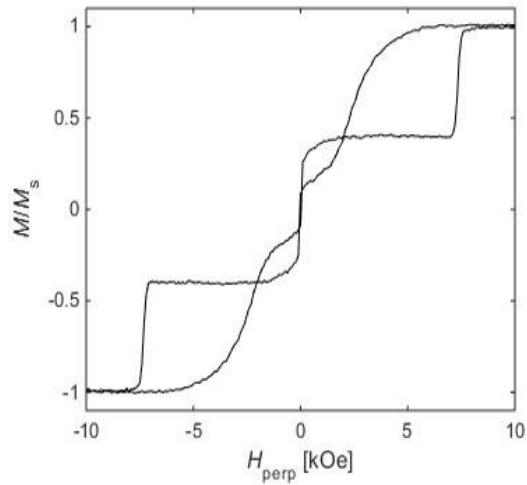

**Supplementary Figure 4| Out-of-plane magnetization curve of the blanket film for the stack structure used in the two-terminal device Ta(3.8)/CoFeB(1)/MgO(1.2)/CoFeB(1.3)/Ta(0.4)/Co(0.4)/Pd(0.6)/Co(0.4)/Ru(0.85)/Co(0.4)/ [Pd(0.6)/Co(0.3)]x3/Ru(1.5).**

### Supplementary Note 4: Ta underlayer etching

Fabrication process for our device is thoroughly described in Method. Here, we show most critical steps in the fabrication.

- Immediate sample transfer to the next vacuum chamber after the etching of magnetic tunnel junction pillar. We found that 10 minutes is enough to oxidize the 3.8-nm-thick Ta underlayer in air. This was not expected because the natural oxidation thickness of Ta is known to be less than 1 nm. We speculate that this could be due to the rough surface of the Ta underlayer because the ion-milling process should increase the surface roughness.

- Estimate of Ta underlayer thickness by 4-point probe resistance measurement of a dummy Ta underlayer wire. This is the most critical step in this fabrication process. Magnetic layer must be fully etched while the thin Ta underlayer must be remained. Supplementary Figure 5 shows the representative data for the resistance of a wide dummy Ta wire as a function of the etching time. Red line corresponds to the resistance of 3.8-nm-thick Ta layer. Etching time is varied from 80% to 120% of the calibrated etching time and resistance of the Ta wire is measured for each sample. The best sample is kept for the following fabrication process. In addition, the etching of the magnetic layer is also confirmed by X-ray photoelectron spectroscopy (XPS). A representative data is shown in Supplementary Fig. 6. Ta peaks are detected while no Fe peak is observed, which confirms a successful removal of the magnetic layer.

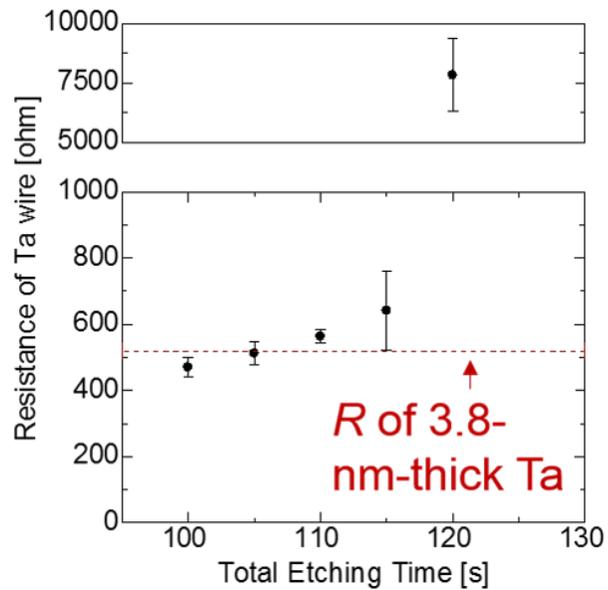

**Supplementary Figure 5| Resistance of the dummy Ta wire as a function of the etching time.**
Red line corresponds to the resistance of 3.8-nm-thick Ta layer.

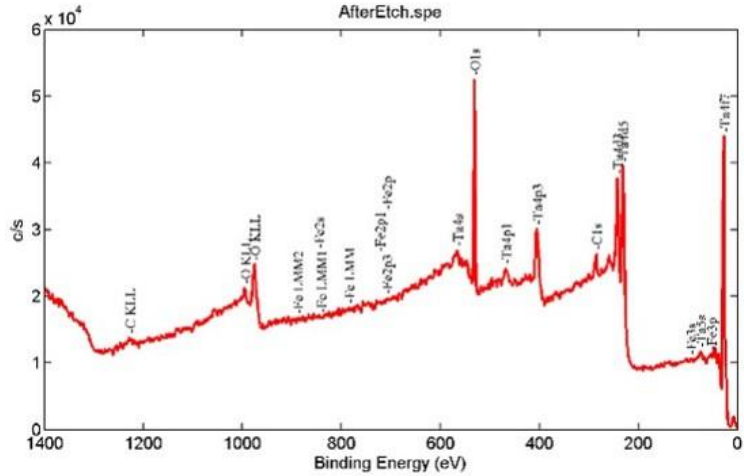

**Supplementary Figure 6| X-ray photoelectron spectroscopy (XPS) spectra of the Ta underlayer.** No Fe peak is observed while Ta peaks are detected, which suggests a successful etching of the magnetic layer.

## Supplementary Note 5: Switching under no magnetic field

Supplementary Figures 7 and 8 show a current-induced magnetic switching hysteresis loop and the critical switching current as a function of pulse length under no external magnetic field for the device with 220-nm-wide underlayer, respectively. We assumed that the switching mechanism is spin transfer torque dominant based on the fact that the P to AP critical current is approximately twice larger than that of AP to P. Thus, the following equation for thermally activated spin-transfer torque switching is used to extract $J_{c0}$:

$$J_\text{c} = J_\text{c0}\left[1 - \frac{1}{\Delta}\ln\left(\frac{\tau}{\tau_0}\right)\right] \tag{5}$$

The extracted intrinsic critical current density $J_{c0}$ is 34.5 MA/cm$^2$ and -19.2 MA/cm$^2$ for P to AP and AP to P switching, respectively. $J_{c0}$ for P to AP switching is almost twice larger than that of P to AP switching. This is in contrast with the switching under external magnetic field and support our assumption that the switching mechanism under no field is spin-transfer torque dominant. It should be also noted that the intrinsic critical current density observed under no field is 5 - 10 times larger than conventional STT-MRAM devices[7,8]. Possible reasons for this discrepancy include:

- Increase of effective damping parameter by the large in-plane current in the Ta underlayer. It has been reported that the effective damping parameter can be much larger than the value measured in the absence of in-plane current[9]. (Spin-transfer torque critical current is inversely proportional to the damping parameter.)

- Weak perpendicular magnetic anisotropy of the reference layer. This leads to a lower spin transfer torque efficiency.

In order to clarify the reason for the large spin-transfer torque critical current density, the critical current density for the device with 20-μm-wide underlayer is measured. In this device, the in-plane current density is approximately 100 times smaller than the device with 220-nm-wide Ta wire. Thus, the in-plane-current-induced torques can be ignored, and we can only consider the contribution from the conventional spin-transfer torque effect by out-of-plane current for this device with 20-μm-wide Ta wire. It should be noted that the stack structure and the process conditions are identical. Supplementary Figure 9 shows the pulse width dependence on the critical current density. Since this switching mechanism must be purely spin-transfer torque, Eq. 5 is utilized to extract the intrinsic critical density. The extracted intrinsic critical current density $J_{c0}$ for the magnetic tunnel junction (MTJ) on the 20-um-wide Ta wire is 16.8 and -8.1 for P to AP and AP to P switching, respectively. These values are almost half of the $J_{c0}$ for MTJ on 220-nm-wide Ta wire. These results support our speculation that the effective damping parameter increases with the in-plane current density in a narrowed Ta wire. In previous studies[10], the increase of the effective damping parameter was not considered.

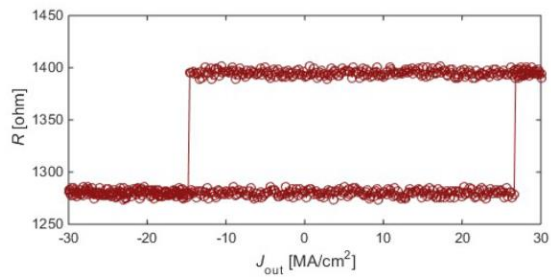

**Supplementary Figure 7| Current induced magnetic switching under no external magnetic field for the device with 220-nm-wide underlayer.**

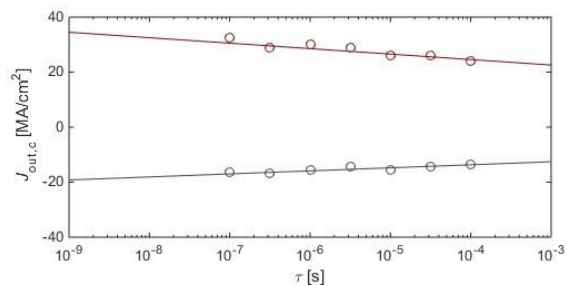

**Supplementary Figure 8| Critical current versus current pulse width under no external magnetic field for the magnetic tunnel junction on 220-nm-wide Ta wire.** (Red line: Parallel to Antiparallel switching. Black line: Antiparallel to Parallel switching). Fitting with Eq. 5 based on thermally activated spin-transfer torque switching.

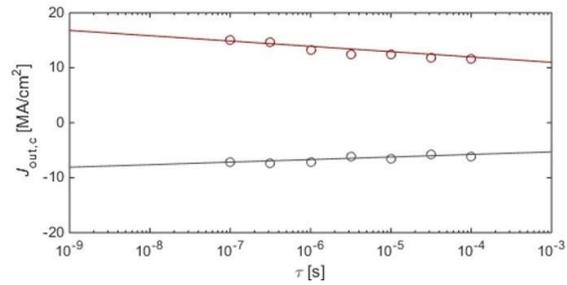

**Supplementary Figure 9| Critical current versus current pulse width under no external magnetic field for the magnetic tunnel junction on 20-μm-wide Ta wire** (Red line: Parallel to Antiparallel switching. Black line: Antiparallel to Parallel switching). Fitting with Eq. 5 based on thermally activated spin-transfer torque switching.

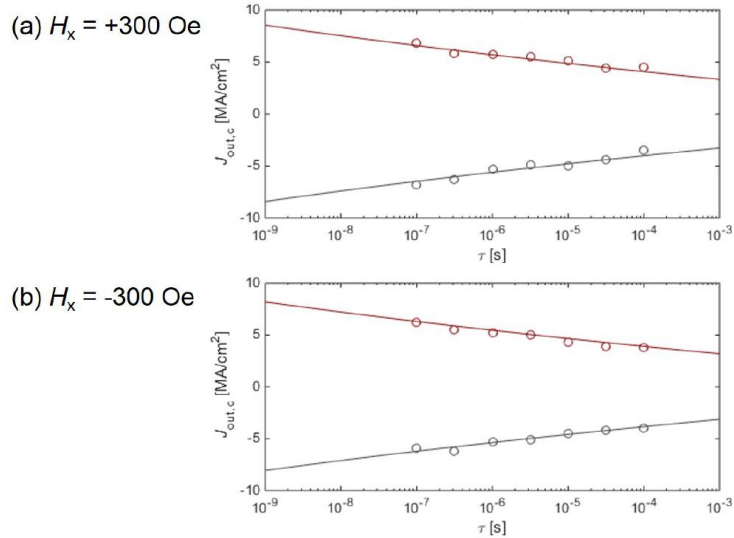

(a) $H_x$ = +300 Oe

(b) $H_x$ = -300 Oe

**Supplementary Figure 10| Critical current versus current pulse width under an in-plane field of 300 Oe for the magnetic tunnel junction on 220-nm-wide Ta wire** (Red line: Parallel to Antiparallel switching. Black line: Antiparallel to Parallel switching). Fitting with Eq. 3 in main text based on thermally activated spin-orbit torque switching.